\documentclass[10pt,english]{IEEEtran}

\usepackage[T1]{fontenc}
\usepackage[latin9]{inputenc}
\usepackage[a4paper]{geometry}
\usepackage{color}
\definecolor{page_backgroundcolor}{rgb}{1, 1, 1}
\pagecolor{page_backgroundcolor}
\usepackage{float}
\usepackage{amsmath}
\usepackage{amssymb}
\usepackage{stmaryrd}
\usepackage{stackrel}

\makeatletter

\providecommand{\tabularnewline}{\\}
\floatstyle{ruled}
\newfloat{algorithm}{tbp}{loa}
\providecommand{\algorithmname}{Algorithm}
\floatname{algorithm}{\protect\algorithmname}

\@ifundefined{lettrine}{\usepackage{lettrine}}{}
\numberwithin{figure}{section}

\@ifundefined{date}{}{\date{}}
\usepackage{babel}
\makeatother

\usepackage{babel}
\begin{document}

\title{NTRUCipher-Lattice Based Secret Key Encryption}

\author{{\normalsize{}{}Maheswara Rao Valluri}{\small{}{}}\\
 {\small{}{}School of Mathematical and Computing Sciences}\\
 {\small{}{}Fiji National University, P.O.Box:7222}\\
 {\small{}{}Derrick Campus, Suva, Fiji Island}\\
 {\small{}{}maheswara.valluri@fnu.ac.fj}}
\maketitle
\begin{abstract}
NTRU cryptosystem has allowed designing a range of cryptographic schemes
due to its flexibility and efficiency. Although NTRU cryptosystem
was introduced nearly two decades ago, it has not yet received any
attention like designing a secret key encryption. In this paper, we
propose a secret key encryption over NTRU lattices, named as NTRUCipher.
This NTRUCipher is designed using modification of the NTRU public
key encryption. We analyze this cipher efficiency and the space complexity
with respect to security aspects, and also show that the NTRUCipher
is secured under the indistinguishability chosen plaintext attack. 

Key words - NTRUCipher; multiple transmission attack; product form
polynomials; secret key encryption.
\end{abstract}

\section{Introduction}

Cryptosystem is classified as the secret key cryptosystem (symmetric
key cryptosystem) and public key cryptosystem (asymmetric cryptosystem)
based on nature of cryptographic key functions and properties. The
secret key cryptosystem could be categorized as the stream cipher
and block cipher based on size of the secret key, plaintext and ciphertext.
The block ciphers are the most influential cryptographic primitives
in designing cryptographic schemes such as encryptions, hash functions,
and message authentication codes. In the secret key encryption, communication
parties, a sender and receiver share and use a common key, as the
secret key, whereas in the public key cryptosystem both the sender
and receiver contain their own secret key and associated public key.
The secret key cryptosystem provides cryptographic services such as
confidentiality, integrity and authentication of a message. The strength
of the secret key encryption relies on two parameters: strength of
algorithm and length of the secret key. The well known secret key
block ciphers are the Advanced Encryption Standard (AES) {[}10{]},
RC5 {[}15{]}, Blowfish {[}3{]}, Data Encryption Standard (DES) {[}2{]},
and International data encryption algorithm (IDEA) {[}17{]}. These
ciphers are built using Feistel network except IDEA for encryption
and decryption. The Feistel network {[}7{]} was designed using XOR
operator and Permutation-Box (P-Box) and Substitution-Box (S-Box).

The proposed cipher presented in this paper is designed using a modification
of the NTRU public key cryptosystem {[}8{]} over the ring, $R=\mathbb{Z}_{q}[x]/(x^{n}-1)$,
where $n$ is a prime and $q$ is a power of 2. The NTRU public key
cryptosystem was first appeared in {[}8{]} due to Hoffstein et al.
in 1998 and it is based on truncated polynomials over finite fields.
Most of the promising candidates of the NTRU cryptosystems are NTRUEncrypt
{[}8{]} and NTRUSign {[}9{]} which are resistance to the Shor's algorithm
{[}14{]} on quantum computers. Damien Stehle et al. proposed a provable
security version of the NTRU public key cryptosystem over the ring,
$R=\mathbb{Z}_{q}[x]/(x^{n}+1)$, where $n$ is a power of 2 and $q$
is a prime, which provides encryption {[}5,6{]} and digital signature
{[}6{]}. There is still scope to work further on their contribution
in terms of setting up a specific security parameters. Recently, Daniel
J Bernstein et al. proposed another variant of the NTRU, named as
NTRU Prime {[}4{]}, over the ring, $R=\mathbb{Z}_{q}[x]/(x^{n}-x-1),$
where $n$ is a prime and $q$ is a power of 2. In this work, we propose
lattice based secret key encryption over NTRU lattices, named as NTRUCipher.
The framework of the encryption and decryption is the same as NTRUEncrypt
{[}8{]} and the decryption is a probabilistic like the NTRUEncrypt
{[}8{]}. Furthermore, we prove the NTRUCipher is secured under the
indistinguishability chosen plaintext attack (IND-CPA), and analyze
efficiency and the space complexity with respect to security aspects. 

The paper is organized as follows: In Section II, we recall the definition
of a secret key encryption and adversary model. Section III presents
truncated polynomial rings. In Section IV, we propose the NTRUCipher
- lattice based secret key encryption and recommend parameters for
the cipher. In Section V, we analyze the NTRUCipher with respect to
performance, security aspects, and space and time complexity. Finally,
we provide conclusion remarks in Section VI.

\section{Preliminaries}

In this Section, we recall the formal definition of a secret key encryption
and its security notations.

\subsection{Notations}

A real valued function $\epsilon(c)<c^{-\lambda}$ is negligible if
for every $\lambda>0$ there exists $c_{\lambda}>0$ such that $\epsilon(c)<c^{-\lambda}$
for all $c>c_{\lambda}$. A probabilistic polynomial time algorithm
is said to be efficient if its running time is polynomial in its input
length. We represent $x$ a random variable sampled from the probability
distribution $D$ as $x\leftarrow D$. The effectiveness of an algorithm
to distinguish between two probability distributions $D_{0}$ and
$D_{1}$ is measured by its distinguishing advantage, defined by $|Pr_{x\leftarrow D_{0}}[A(x)=1]-Pr_{x\leftarrow D_{1}}[A(x)=1]|$.
We say that a decision problem is hard if there does not exist an
efficient algorithm for it that has a non-negligible advantage in
$\lambda$. The statistical distance $\triangle(D_{0};D_{1})$ between
two distributions $D_{0},D_{1}$ on some countable domain $X$ is
defined as $\triangle(D_{0};D_{1})=|D_{0}(x)-D_{1}(x)|$.

\subsection{Secret Key Encryption}

The goal of the secret key encryption is to furnish confidentiality
of communications of two or more parties. In the secret key encryption,
a common secret key is shared among the communication parties, before
decryption of the ciphertext, to furnish confidentiality of the plaintext.
The following definitions are acquired from {[}12,13{]}. For further
details, the reader is referred to {[}12,13{]}.

\textit{\textcolor{black}{Definition}} 1. A secret key encryption
$\mathcal{SE}=(\mathcal{K}eyGen,\mathcal{E}nc,\mathcal{D}ec)$ consists
of the following three algorithms:

\subsubsection*{(i) Key Generation}

Key Generation $(\mathcal{K}eyGen)$ is a randomized algorithm that
outputs a random key. When the algorithm is run, a different key is
generated every time. Note that in this case, the input for the algorithm
is a null. The randomized output secret key is represented with a
dollar symbol \$:

1: Input: null 

2: $k\xleftarrow{\$}KeyGen(1^{\lambda})$

\subsubsection*{(ii) Encryption }

Encryption ($\mathcal{E}nc)$ is a randomized algorithm that takes
a plaintext $\mu$ and the secret key $k$ as input, and outputs a
ciphertext $c$.

1: Input: $\mu\in\mathcal{D_{\mu}}$, $k\in\mathcal{D}_{k}$, where
$D_{\mu}$ is the plaintext space, $\mathcal{D}_{k}$ is the secret
key space and secret key $k$ is generated by the $\mathcal{K}eyGen(1^{\lambda})$
algorithm.

2: $c\xleftarrow{\$}\mathcal{E}nc(\mu,k)$, otherwise $\varPhi$ if
$\mu\notin\mathcal{D}_{\mu}$) $\forall c\in\mathcal{D}_{c}$, where
$\mathcal{D}_{c}$ is the ciphertext space.

\subsubsection*{(iii) Decryption }

Decryption $(\mathcal{D}ec)$ is a deterministic algorithm.

1: Input: $c\in\mathcal{D}_{c}$, $k\in\mathcal{D}_{k}$, where $\mathcal{D}_{c}$
is the ciphertext space and $k$ is the secret key, generated by the
$KeyGen(1^{\lambda})$ algorithm.

2: $\mu\leftarrow\mathcal{D}ec(c,k)$ $\forall c\in\mathcal{D}_{c}$.

Correctness of the secret key encryption works as follows: We say
that a secret key encryption $\mathcal{SE}=(\mathcal{K}eyGen,\mathcal{E}nc,\mathcal{D}ec)$
is correct, if it holds for every plaintext $\mu$ that $Pr[Dec_{k}(Enc_{k}(\mu))\neq\mu:k\xleftarrow{\$}KeyGen(1^{\lambda})]<negl(\lambda).$

\subsection{Security Notations}

The aim of an adversary $\mathcal{A}$ is to capture the secret key
of the secret key encryption and then perceive the plaintext corresponding
to the ciphertext. We assume that the adversary does not have prior
knowledge of the secret key $k.$ The indistinguishability under chosen-plaintext
attack or IND-CPA security is defined as follows:\textbf{ }

\textit{\textcolor{black}{Definition}} 2. Let $\mathbb{\mathcal{A}}$
be an adversary. Let $\mathcal{SE}=(\mathcal{K}eyGen,\mathcal{E}nc,\mathcal{D}ec)$
be the secret key encryption. Let us define the following experiment
between a challenger and $\mathcal{A}$: 

\textit{Experiment: IND-CPA$_{\mathcal{SE}}^{b}(\mathcal{A})$}

1: The challenger runs $k\xleftarrow{\$}\mathcal{K}eyGen(1^{\lambda})$.

2: $\mathcal{A}$ outputs a pair of plaintexts $(\mu_{0},\mu_{1})$
of the same length and sends $(\mu_{0},\mu_{1})$ to the challenger.

3: The challenger computes $c^{*}\xleftarrow{\$}\mathcal{E}nc(\mu_{b},k)$
and then sends $c^{*}$ to the adversary.

4: $\mathcal{A}$ continues its computation and outputs : $b^{'}$.

5: Output 1 if $b=b^{'}$, and 0 otherwise. 

\textit{\textcolor{black}{Definition}} 3. A secret key encryption
$\mathcal{SE}=(\mathcal{K}eyGen,\mathcal{E}nc,\mathcal{D}ec)$ is
indistinguishable under chosen plaintext attack, if it holds for all
probabilistic polynomial time adversary $\mathcal{A}$ that

\begin{flushleft}
$Adv_{\mathcal{SE}}^{IND-CPA}(\mathcal{A})=|Pr[Exp{}_{\mathcal{SE}}^{IND-CPA}(\mathcal{A})=1]-\frac{1}{2}|\leq negl(\lambda)$.
\par\end{flushleft}

\section{Truncated Polynomial Rings }

Let $q\in\mathbb{N}$ be a prime. We write $\mathbb{Z}_{q}$ for the
integer modulo $q$ and represents this set by integers in the range
$(-\frac{q}{2},\frac{q}{2})$. The truncated polynomial ring $R_{n,q}=\mathbb{Z}_{q}[x]/(x^{n}+1)$
consists of all polynomials with coefficients in $\mathbb{Z}_{q}$
and degree less than $n$. An element $f\in R_{q}$ is represented
as a polynomial,

$f=\stackrel[i=0]{n-1}{\sum}f_{i}x^{i}=[f_{0},f_{1},.....,f_{n-1}]$.

Two polynomials $f,g\in R_{q}$ are multiplied by the ordinary convolution,

$(f*g)_{k}=\stackrel[i+j\equiv k(mod\,n)]{}{\sum}(f_{i}.g_{j}),k=0,1,...,n-1,$
which is commutative and associative. The convolution product is represented
by {*} to distinguish it from the multiplication in $\mathbb{Z}_{q}$.
We define a center $l_{2}-$norm of an element $f\in R_{q}$ by $\Vert f\Vert_{2}=(\stackrel[i=0]{N-1}{\sum}f_{i}-\bar{f})^{1/2}$,
where $\bar{f}=\frac{1}{N}\stackrel[i=0]{N-1}{\sum}f_{i}$ and the
infinity norm is $\Vert f\Vert_{\infty}=$max$_{0\leq i\leq n-1}\bigparallel f_{i}\bigparallel.$

\textit{Lemma 1{[}14,15{]}.} For any $f,g\in R_{n,q}=\mathbb{Z}_{q}[x]/(x^{n}+1)$,
$\bigparallel f.g\bigparallel\leq\sqrt{n}.\bigparallel f\bigparallel.\bigparallel g\bigparallel$
and $\bigparallel f.g\bigparallel_{\infty}\leq n.\bigparallel f\bigparallel_{\infty}.\bigparallel g\bigparallel_{\infty}.$

\subsection{Cryptographic Assumptions}

In this Subsection, we define the NTRUCipher ciphertext cracking problem
for which the parameters are chosen as recommended in the table 1.
The search and decision ciphertext cracking problems are defined as
follows:

\subsubsection{\textit{Search NTRUCIpher Ciphertext Cracking Problem}}

Given $c=r*k^{-1}+\mu(mod\,q)\in R_{n,q}$ , with $r\stackrel{\$}{\leftarrow}\mathbb{P}_{n}(a_{1},a_{2},a_{3}),k\stackrel{\$}{\leftarrow}\mathbb{P}_{n}(a_{1},a_{2},a_{3})$
, compute $(r,k,\mu)$.

\subsubsection{\textit{Decision NTRUCIpher Ciphertext Cracking Problem}}

Given $c=r*k^{-1}(mod\,q)\in R_{n,q}$, distinguish whether $c$ is
sampled from the distribution $D_{0}=\{c=r*k^{-1}(mod\,q):r\stackrel{\$}{\leftarrow}\mathbb{P}_{n}(a_{1},a_{2},a_{3}),k\stackrel{\$}{\leftarrow}\mathbb{P}_{n}(a_{1},a_{2},a_{3})\}$
or from the uniform distribution $D_{1}=U(R_{n,q})$.

We assume that the decision NTRUCipher Ciphertext Cracking Problem
is hard to indistingush computationally.

\section{NTRUCipher}

In this Section, we propose NTRUCipher which is drawn from NTRUEncrypt
{[}2,4{]} by modification. In this cipher, we use a ring of $R_{n,q}=\mathbb{Z}_{q}[x]/(x^{n}+1)$
and propose NTRUCipher-lattice based secret key encryption.

\subsubsection*{System Parameters}

The cipher would have three integer parameters, $n,p$ and $q$. The
integer $n$ has to be $2^{l}$ , $p$ is a small prime, and $q$
is large prime such that $gcd(p,q)=1$ , and $p<q$. 

\begin{table}[H]
\caption{NTRUCipher - System Parameters}

\centering{}%
\begin{tabular}{|ll|}
\hline 
1. & $n$- Degree Parameter\tabularnewline
2. & $q$ - Large Modulus\tabularnewline
3. & Ring Parameters, $R_{n,q}=\frac{\mathbb{Z}_{q}[x]}{(x^{n}+1)}$\tabularnewline
4. & $p$ - Plaintext space modulus\tabularnewline
5. & $a_{1},a_{2},a_{3}$ - Non-zero coefficient counts for \tabularnewline
 & product form polynomial terms.\tabularnewline
6. & $\mu$ - Plaintext .\tabularnewline
7. & $c$ - Ciphertext\tabularnewline
8. & $k$ - Secret key \tabularnewline
9. & r - Ephemeral key\tabularnewline
10. & $\mathcal{D}_{\mu}$ - Plaintext space\tabularnewline
11. & $\mathcal{D}_{c}$ - Ciphertext space\tabularnewline
12. & $\mathcal{D}_{k}$- Secret key space\tabularnewline
13. & $\mathcal{D}_{r}$- Ephemeral key space\tabularnewline
14. & $\mathbb{B}_{n}=$\{Binary Polynomials\}\tabularnewline
15. & $\mathbb{T}_{n}=$\{Ternary Polynomials\}\tabularnewline
16. & $\mathbb{T}_{n}(a,e)=$\{Ternary polynomials with exactly \tabularnewline
 & $a$ ones and $e$ minus ones\}\tabularnewline
17. & $\mathbb{P}_{n}(a_{1},a_{2},a_{3})=$\{Product form of \tabularnewline
 & polynomials $A_{1}*A_{2}+A_{3}:A_{i}\in\mathbb{T}_{n}(a_{i},a_{i})\}$\tabularnewline
\hline 
\end{tabular}
\end{table}

\subsubsection*{(i) Key Generation}

1. The secret key $k$ is a polynomial of the form $p.k^{'}+1$, where
$k^{'}$ is generated by product form of polynomials, $\mathbb{P}_{n}(a_{1},a_{2},a_{3})$.
Note that this form ensures that $k$ has inverse $1$ modulo $p$. 

\begin{algorithm}[H]
\caption{NTRUCipher-Key Generation}

\textbf{Input:} A set of system parameters

1: \textbf{repeat}

2: $k^{'}\stackrel{\$}{\leftarrow}\mathbb{P}_{n}(a_{1},a_{2},a_{3})$

3: $k=1+p.k^{'}\in R_{n,q}$

4: until $k$ is invertible in $R_{n,q}$

\textbf{Output :} Secret key $k.$
\end{algorithm}

\subsubsection*{(ii) Encryption }
\begin{enumerate}
\item To encrypt a plaintext $\mu\in\{-(p-1)/2,..,+(p-1)/2\}^{n}$ with
the secret key $k$, first a polynomial $r$ is randomly sampled in
$\mathbb{P}_{n}(a_{1},a_{2},a_{3})$ such that $r*k^{-1}(mod\,q)\in$
$R_{n,q}$.
\item Compute ciphertext $c=(p.r*k^{-1}+\mu)(mod\,q)$.
\end{enumerate}
\begin{algorithm}[H]
\caption{NTRUCipher-Encryption }

\textbf{Input:} Secret key $k$, message $\mu\in\{-(p-1)/2,..,+(p-1)/2\}^{n}$,
and set a parameter, $p=3$.

1: \textbf{repeat}

2: $r\stackrel{\$}{\leftarrow}\mathbb{P}_{n}(a_{1},a_{2},a_{3})$

3: $c=(p.r*k^{-1}+\mu)(mod\,q)\in R_{n,q}$

\textbf{Output :} Ciphertext $c$
\end{algorithm}

\subsubsection*{(iii) Decryption }

To decrypt the ciphertext $c$ with respect to the secret key $k$,
\begin{enumerate}
\item Compute first ciphertext, $c^{'}=c*k(mod\,q)$, and center the coefficient
of $c^{'}$ in $(-q/2,q/2)$. 
\item Then, compute $\mu^{'}=c^{'}(mod\,p)$, and center the coefficient
in $(-p/2,p/2)$ to get the plaintext $\mu$.
\end{enumerate}
\begin{algorithm}[H]
\caption{NTRUCipher-Decryption }

\textbf{Input:} Secret key $k$, Ciphertext $c\in R_{n,q}$, and set
a parameter, $p=3$.

1: $c^{'}=c*k(mod\,q)\in R_{n,q}$ 

2: Center the coefficients of $c^{'}$ in $(-q/2,q/2)$

3: $\mu^{'}=c^{'}(mod\,p)$

4: Center the coefficients of $\mu^{'}$ in $(-p/2,p/2)$

5: Result $=\mu$

\textbf{Output :} Plaintext $\mu$
\end{algorithm}

\subsubsection*{Completeness}

To decrypt the ciphertext $c$ with respect to the secret key $k$,
\begin{enumerate}
\item Compute first cipher text $c^{'}=c*k(mod\,q)=(p.r+\mu*k)(mod\,q)$,
and center the coefficient of $c^{'}$ in $(-q/2,q/2)$. 
\item Then, compute $\mu^{'}=c^{'}(mod\,p)=[p.(r+\mu*k^{'})+\mu](mod\,p)$
, and center the coefficient in $(-p/2,p/2)$ to get the plaintext
$\mu$.
\end{enumerate}

\subsection{Probability of decryption failure}

In this subsection, we estimate decryption failure of the NTRUCipher
in terms of probability. For a successful decryption of the ciphertext
$c$ for the plaintext $\mu$ using the given secret key $k$, the
coefficient of the $c^{'}=[p.(r+\mu*k^{'})+\mu]$ must be in the range
of less than $\frac{q}{2}.$ By the triangle inequality, the following
relation holds:

$\bigparallel c^{'}\bigparallel_{\infty}\leqslant p.(\bigparallel r\bigparallel_{1}+\bigparallel\mu\bigparallel_{\infty}\bigparallel k^{'}\bigparallel_{1})+1]$
. 

We assume that $r$ and $k^{'}$ are chosen in the product form and
the plaintext $\mu$ is a ternary polynomial. Note that the decryption
failure can be avoided by ensuring $q>8p(2a_{1}a_{2}+a_{3})+2$ .
We hereby set a probabilistic bound to estimate the probability; 

Prob(a given coefficients of $r+\mu*k^{1}$ has absolute value $\geq$$\mathit{B}$).

We choose $r$ and $k^{'}$ in the product form such that $r=r_{1}*r_{2}+r_{3}$,
$k^{'}=k_{1}^{'}*k_{2}^{'}+k_{3}^{'},$ where each $r_{i}$ and $k_{i}^{'}$
has exactly $a_{i}$ coefficients equal to 1, $a_{i}$ coefficients
equal to -1 and the rest of the coefficients equal to 0. When the
coefficients of the plaintext $\mu$ are chosen from $\{-(p-1)/2,..,+(p-1)/2\}^{n}$,
the probability of taking $0$ as coefficients of plaintext $\mu$
is $\frac{a_{\mu}}{n}$, and taking $\pm1$ as coefficients is $(1-\frac{a_{\mu}}{n})$.
The coefficients of $r+\mu*k^{1}$ are expected to be distributed
according to the convolution of normal distribution with standard
deviation, $\sigma=\sqrt{(4a_{1}a_{2}+2a_{3})(2-\frac{a_{\mu}}{n})}$,
which is computed by adopting technique of section 6 of {[}11{]}. 

The probability that a normally distributed random variable with mean
0 and standard deviation $\sigma$ exceeds $B$ in absolute value
is given by the complementary error function, $erfc(B/(\sqrt{2}$$\sigma$).
Thus, the probability that any of the $n$ coefficients of $r+\mu*k^{'}$
is greater than $B$ is bounded by $n.erfc(\mathit{B}/(\sqrt{2}$$\sigma)$.
With respect to security parameter $\lambda$ this imposes the constraint
$n.erfc((q-2)/(2\sqrt{2}p\sigma)<2^{-\lambda}$, where $\sigma=\sigma(n,a_{1},a_{2},a_{3},a_{\mu}).$

\subsection{Parameter Sets and Sample space}

In this Subsection, we give set of parameters for the NTRUCipher.
We discuss how the NTRUCipher parameters are chosen.

\subsubsection{Binary polynomials}

Binary polynomials $\mathbb{B}_{n}$ are used in this cipher to generate
product form of polynomials. These can be easy to implement in software
and hardware. A disadvantage is that binary polynomials are by definition
unbalanced. Therefore, when $k(1)\neq0$, as a consequence information
on the plaintext $\mu$, namely$\mu(1)$ leaks.

\subsubsection{Ternary polynomials}

We define $\mathbb{T}_{n}$ as the set of all ternary polynomials,
a particular case $\mathbb{T}_{n}(a,e)$ with $a$ coefficients are
$1$, $e$ coefficients are $-1$ and rest of the coefficients are
$0$. These ternary polynomials are used to make product form of polynomials.

\subsubsection{Product form of polynomials}

Product form of polynomials $\mathbb{P}_{n}$ are generated by $\mathbb{P}_{n}(a_{1},a_{2},a_{3})=\{a_{1}*a_{2}+a_{3}:a_{1}\in\mathbb{T}_{n}(a,e),a_{2}\in\mathbb{T}_{n}(a,e),a_{3}\in\mathbb{T}_{n}(a,e)$\}.
These polynomials are used in this cipher to choose the secret key
$k$ and ephemeral key $r$. For instance, the secret key $k$ is
chosen of the form $k=1+p.k^{'}$, where $k^{'}\in\mathbb{P}_{n}(a_{1},a_{2},a_{3})$.
The number of non-zero coefficients in $k^{'}$ and $r$ are crucial
for the performance of the encryption. Note that convolution can be
faster if there are a small number of non-zero elements in the polynomial.
An advantage of the product form of polynomials is that they allow
for exceptionally fast convolution without Fourier transforms.

\subsubsection{Secret Key Space}

The space of secret key $\mathcal{D}_{k}$ consists of all polynomials
that are derived from $\mathbb{P}_{n}(a_{1},a_{2},a_{3})$. If $n$
and $q$ are fixed in advance, we can choose $a_{1},a_{2}$ and $a_{3}$
with $\pm1$'s for the secret key $k$. When we take $a_{1}\approx a_{2}\approx a_{3}$,
the expected number of non-zero coefficients in $k$ are $4a_{1}a_{2}+2a_{3}\approx\frac{2n}{3}$
which is an optimal for key selection. The space complexity of the
secret key is $O((2\bigparallel k\bigparallel_{\infty}+1)^{n}).$
The size of the secret key is $n\lfloor log_{2}(2\bigparallel k\bigparallel_{\infty}+1)\rceil$.

\subsubsection{Ephemeral key space}

The space of ephemeral key $\mathcal{D}_{r}$ consists of all polynomials
that are derived from $\mathbb{P}_{n}(a_{1},a_{2},a_{3})$. The space
complexity of the ephemeral key is $O((2\bigparallel r\bigparallel_{\infty}+1)^{n}).$
The size of the ephemeral key is $n\lfloor log_{2}(2\bigparallel r\bigparallel_{\infty}+1)\rceil$.

\subsubsection{Plaintext space}

The plaintext space $\mathcal{D}_{\mu}$ is defined as $\mathcal{D}_{\mu}=\{\mu\in R_{n,q}/\mu$
has coefficients in $(-\frac{p-1}{2},\frac{p-1}{2})\}$, assuming
$p$ is odd prime. The recommended parameter, $p=3.$ The space complexity
of the plaintext space is $O(p^{n})$. The size of the plaintext is
$n\lfloor log_{2}p\rceil$.

\subsubsection{Ciphertext space}

The ciphertext space $\mathcal{D}_{c}$ is defined as $\mathcal{D}_{c}=\{c\in R_{n,q}/c$
has coefficients in $(-\frac{q-1}{2},\frac{q-1}{2})$, assuming $q$
is odd prime. The complexity of the ciphertext space is $O(q^{n})$.
The size of the ciphertext is $n\lfloor log_{2}q\rceil.$

\subsection{Efficiency.}

We provide efficiency of the NTRUCipher in terms of $O$ notations.
In this NTRUCipher, we can perform addition and multiplication of
two polynomials over the ring, $R_{n,q}=\mathbb{Z}_{q}[x]/(x^{n}+1)$
in $O(n)$ and $O(n^{2}\lfloor log_{2}q\rceil)$ bits operations respectively.
Thus, it follows that the cost of encryption and decryption in both
cases is $O(n^{2}\lfloor log_{2}q\rceil)$ bit operations.

\subsection{Concrete Parameters Set}

Our recommendations for parameters of the NTRUCipher suggests taking
$n=256,$ $p=3,$ $q=1087$ , $a_{1}=5,a_{2}=5,a_{3}=5$, and $a_{\mu}=102$
so that we will ensure upper bound on the security parameter that
our decryption failure probability is less than $2^{-80}$.

\section{Security Analysis}

\subsubsection{Brute-force attack}

Brute-force attack is one of the generic cryptographic attacks in
which one could try for every possible key permutation until it finds
the secret key. The feasibility to find out the secret key in brute
force attack relies on key space. In this cipher, the key space order
for $n$ bit length is $O((2\bigparallel k\bigparallel_{\infty}+1)^{n}).$
In the proposed cipher, we use minimum key length of 256 bits. Therefore,
one should try $(2\bigparallel k\bigparallel_{\infty}+1)^{256}$ bit
permutations to find the secret key which is large enough for brute
force attack. The plaintext $\mu$ is chosen as a polynomial with
ternary coefficients of degree 256 in the NTRUCipher, one should try
$3^{256}$ bit permutations to find the plaintext $\mu$ on the brute
force attack. According to knowledge of the author, currently complexity
order $2^{80}$ is considered as the lower bound of security for brute
force attack.

\subsubsection{Multiple transmission attack}

In this NTRUCipher, one would send a single plaintext $\mu$ multiple
times using the same secret key $k$ and different ephemeral keys,
$r$'s. In this scenario, one would transmit $t$ ciphertexts such
that $c_{i}=p.r_{i}*k^{-1}+\mu(mod\,q)$ for $i=1,2,3....,t.$ The
adversary can then compute $(c_{i}-c_{1})=(r_{i}-r_{1})*k^{-1}(mod\,q)$.
Note that $c=r*k^{-1}(mod\,q)$ , where $c=c_{i}-c_{1}$ and $r=r_{i}-r_{1}$,
has same structure as the public key generation of the NTRUEncrypt
{[}2,4{]}. Since $c=r*k^{-1}(mod\,q)$, there exists $u\in R_{n,q}$
with $u=\frac{-k*c+r}{q}$ such that $[k,u].\mathfrak{L}_{c}=[k,r]$,
where $\mathfrak{L}_{c}=\left[\begin{array}{cc}
I & c\\
0 & qI
\end{array}\right]$. By choosing the appropriate parameters for $[k,r]$, one can find
the secret key $k$ by solving SVP (approximate SVP) in $\mathfrak{L}_{c}$.

\subsubsection{Chosen Plaintext attack}

In chosen plaintext attack, the adversary can randomly choose plaintexts
to be encrypted and gets corresponding ciphertexts. The goal of the
attack is to get access to the information that reduces the security
of the secret key encryption. We hereby prove the NTRUCipher is indistinguishability
chosen plaintext attack-secure. We use the following game IND-CPA
between a challenger and an adversary.

\textit{Theorem 1. }The NTRUCipher-Lattice based secret key encryption
is INA-CPA Secure if the decision NTRUCipher ciphertext cracking problem
is hard.

\textit{Proof:} Let $\mathcal{A}$ be an adversary. $\mathcal{A}$
is given random oracle access to $NTRUCipher-Enc(.)$ and outputs
two plaintexts $\mu_{0},\mu_{1}$ of equal length of $n$ and sends
to a challenger. The challenger runs an algorithm $\boldsymbol{C}$
and picks $b\leftarrow\{0,1\}$, and computes challenging ciphertext
$c^{*}=(p.r^{*}*k^{*-1}+\mu_{b})(mod\,q)$ instead of $c=(p.r*k^{-1}+\mu)(mod\,q)$.
Then, the challenger sends the challenging ciphertext $c^{*}$ to
the adversary $\mathcal{A}$. Eventually, when the adversary $\mathcal{A}$
outputs his guess $b^{'}$ for $b$ , the algorithm $\boldsymbol{C}$
outputs 1 if $b^{'}=b$, and 0 otherwise. 

We now compute advantage function of the adversary under indistinguishability
(IND) chosen plaintext attack(CPA) experiment: $Adv_{\mathcal{NTRU}Cipher}^{IND-CPA}(\mathcal{A}):=|Pr[Exp_{\mathcal{NTRU}Cipher}^{IND-CPA}(A)=1]-\frac{1}{2}|$.
We say that the NTRUCipher satisfies IND-CPA, if the advantage $Adv_{\mathcal{NTRU}Cipher}^{IND-CPA}(\mathcal{A})$
is negligible for any polynomial time adversary $\mathcal{A}$. 

If we fix $k^{*}\in R_{n,q}$, the encrypted oracle uses the value
$r^{*}$ to answer at least one of the $n$ queries of $\mathcal{A}$.
Let $q_{n}$ denote the total number of queries made by $\mathcal{A}$.
In this experiment, $\mbox{\ensuremath{\mathcal{A}}}$ can succeed
by one of the two possibilities. If $r^{*}$ is drawn from $D_{0}$
for some queries of $\mathcal{A}$, then $\mathcal{A}$ succeeds and
the algorithm $\boldsymbol{C}$ returns 1 with probability is at most
$\frac{q_{e}}{(2\bigparallel r^{*}\bigparallel_{\infty}+1)^{n}}$.
If $r^{*}$ is drawn from $D_{1}$ for some queries of $\mathcal{A}$,
then $\mathcal{A}$ succeeds only by guessing $b$, and the algorithm
$\boldsymbol{C}$ returns 1 with probability $\frac{1}{2}$. Thus,
we have $Pr[Exp_{\mathcal{NTRU}Cipher}^{IND-CPA}(A)=1]\leq\frac{1}{2}+\frac{q_{e}}{(2\bigparallel r^{*}\bigparallel_{\infty}+1)^{n}}$.
The advantage of the adversary of the NTRUCipher, $Adv_{\mathcal{NTRU}Cipher}^{IND-CPA}(\mathcal{A}):=|Pr[Exp_{\mathcal{NTRU}Cipher}^{IND-CPA}(A)=1]-\frac{1}{2}|$$\leq\frac{1}{2}+\frac{q_{e}}{(2\bigparallel r^{*}\bigparallel_{\infty}+1)^{n}}-\frac{1}{2}\leq\frac{q_{e}}{(2\bigparallel r^{*}\bigparallel_{\infty}+1)^{n}}\leq negl(\lambda)$.
Therefore, the NTRUCipher is IND-CPA secure.

\section{Conclusion}

In this paper, we have proposed a secret key encryption which is based
on truncated polynomials over NTRU lattices instead of classical well
known Feistel structure {[}7{]}. Attacks such as brute force attack,
multiple transmission attack and IND-CPA have been exposed against
the proposed NTRUCipher. We have also recommended a set of parameters
and analyzed security aspects and efficiency. A disadvantage of the
NTRUCipher is that size of ciphertext is very larger compared to other
existing secret key block ciphers {[}2,10{]}. Further work is required
for designing the NTRUCipher based homomorphic secret key encryption
and message authentication code.

\end{document}